# A new method to determine the mean density of massive Solar System bodies

Konstantin P. Kobzar

*Novosibirsk State Pedagogical University, Novosibirsk, 630126, Russia*

**ABSTRACT**

Mean densities of major and dwarf planets are possible to calculate by the values of the planets' distance to the Sun, the mean densities of massive natural satellites of planets are computable by the satellite's distance to the Sun and the primary. The article hypothesizes that the mean density of a body was affected by the gravitational field during the body formation in the formation point, and the gravity was influenced by the Sun and a hypothetical supermassive belt in the region beyond Neptune, and by the primary also, in case of the natural satellites. The mean densities obtained by the traditional methods and through the newly proposed approach characterize different life stages of celestial bodies, and the comparative analysis of these mean density values can be a useful tool in studying migration of the bodies in the Solar System and in other planetary systems.

___________________________

E-mail: cenat2000@mail.ru

## 1. Introduction

Harmony features the structure of the Solar System. This is proved by Kepler's laws that describe regular geometry of orbits and the ordered coordination of celestial bodies in space and in time. Major planets move in the same direction, in approximately the same plane, and some other celestial objects behave in the same manner. The harmony is contributed to by orbital resonances and by spin-orbital resonances. With that, these well-known relationships only characterize mutual position and motion of the celestial bodies, while changes in their physical characteristics, which are far poorly studied, could not be judged accidental in the balanced and coordinated universe. Any body has mass, size, shape, as well as magnetic and other properties. One of the properties of a body is its mean density, which is considered as a source of information and is used in determining compositions and structures of objects.

Over several decades, many researchers have been studying mean densities of celestial bodies. The enormous work has resulted in the gravitational and non-gravitational methods to estimate the mean densities, as well as in numerous estimates and analyses of the mean densities of a variety of objects: major planets, dwarf planets, natural satellites, and small Solar System bodies (SSSBs), inclusive of asteroids, trans-Neptunian objects and cometary nuclei. The present article uses generally the recently obtained estimates of the mean densities of the Solar System bodies.

Here we tried to use the most recent data. Eight major planets are grouped as 4 terrestrial planets with relatively high values of their mean densities (Strom, 2007; Weissman, 2007), and 4 gas giant planets with low mean densities (Jacobson et al., 2006; Marley and Fortney, 2007). Currently there are five dwarf planets, among which Ceres lies in the main asteroid belt, and the rest dwarf planets are TNOs. Baer et al. (2011) determined the mean density of Ceres; Tholen et al. (2008) – Pluto; Levi and Podolak (2011) – Haumea; Sicardy et al. (2011) – Eris. The mean density of the fifth dwarf planet Makemake has not yet been determined.

The mean densities of seven TNOs have also been analyzed. Lacerda and Jewitt (2007) determined the mean density of Varuna; Fraser and Brown (2010) – Quaoar; Grundy et al. (2007) – Ceto; Brown et al. (2010) – Orcus; Stansberry et al. (2006) - 1999 TC36; Dotto et al. (2008) - 2000 GN171; Orly and Re'em (2010) - 2001 QG298.

Asteroids constitute more than a half amount of objects with the known mean densities. Baer et al. (2011) calculated the mean densities for asteroids, numbers 2-4, 6-11, 13-21, 29, 31, 39, 47-49, 52, 65, 87, 88, 90, 121, 130, 243, 253, 283, 379, 444, 451, 511, 704, 804, 185851, 276049, 2000 UG11. Fienga et al. (2009) determined the mean densities of twelve asteroids, numbers 12, 23, 41, 89, 128, 129, 139, 173, 192, 354, 409, 532. The mean densities of a few asteroids were estimated in the works by Marchis et al. (2008a), numbers 22, 45, 107, 762; Baer and Chesley (2008), number 24; Chesley et. al. (2010), number 189; Descamps et al. (2011), number 216; Yeomans et al. (2000), number 433; Mueller et al. (2010), number 617; Rojo and Margot (2011), number 702; Marchis et al. (2008b), number 3749; Abe et al. (2006), number 25143; Ostro et al. (2006), number 66391; Brooks (2007), number 164121.

Sosa and Fernández (2009) determined the mean densities for nuclei of comets Halley, Encke, d'Arrest, Tempel 1, Tempel 2, Borrelly, Kopff, Honda-Mrkos-Pajdusakova, Wirtanen, Churyumov-Gerasimenko, Wild 2.

The data on the mean densities are available for three satellites of terrestrial planets, for Pluto's satellite, and for 27 satellites of large planets. Wieczorek et al. (2006) high-accuracy calculated the Moon's mean density; Andert et al. (2010) produced the mean density estimate for Phobos; Smith et al. (1995) – Deimos; Person et al. (2006) – Charon. The mean densities of 5 Jupiter's satellites are known. Lopes (2007) determined the mean density of Io; Greenberg (2010) - Europa, Ganymede, Callisto, Amalthea. The mean densities of 16 Saturnian moons are known. Thomas (2010) determined the mean densities of Mimas, Enceladus, Tethys, Dione, Rhea, Hyperion, Iapetus, Phoebe, Janus, Epimetheus, Atlas, Prometheus, Pandora, Pan, Daphnis; Jacobson et al. (2006) – Titan. Jacobson et al. (1992) determined the mean densities of Uranian satellites Miranda, Ariel, Umbriel, Titania, Oberon; Person et al. (2006) – Neptune's satellite Triton. Thus and so, the present article analysis includes data on the known mean densities of 131 objects of the Solar System. The article treats a body's mean density as a principal and self-sufficient, rather than subsidiary, characteristic of the Solar System bodies.

## 2. Gravitational field and the mean densities of bodies in the Solar System

This study is aimed at analyzing spatial distribution of mean densities of celestial bodies in the Solar System. Density is defined as mass per unit volume; mean density is the object's mass divided by volume. Mass, as a physical value, determines gravity of bodies; therefore, mass and, hence, density relate to the gravitational field. According to present-day beliefs, bodies of the Solar System formed under the dominant gravitational influence. That is, on the one hand, every macro- or microscopic body's mass affected the gravitational field; on the other hand, the gravitational fields of existent bodies had influence on the forming bodies, including their densities. Inasmuch as the gravitational field in the Solar System was nonuniform, the nonuniformity could show itself in discrepancy of the mean densities of the forming bodies.

The mean densities of more than 130 bodies in the Solar System are determined to date. These bodies are the Sun, major planets, dwarf planets, moons of the major and dwarf planets, comets, main belt asteroids, Kuiper belt and other trans-Neptunian objects (TNOs). Unlike other physical properties, for instance, a mass or a volume, the determined mean densities of the Solar System bodies have relatively close value range, within one-two orders of magnitude. The lowermost values of densities belong to nuclei of comets; the uppermost values belong to terrestrial planets and a few asteroids. Spatial distribution of the mean densities is reasonable to consider relative to the Solar System barycenter, or relative to the Sun, which is almost one and the same. The present study used semimajor axis of orbit of celestial bodies as a spatial parameter. Orbital parameters of the all objects taken from JPL http://ssd.jpl.nasa.gov/

Though the bodies in the Solar System always had and have now the gravitational effect on each other, this effect is possible to neglect, as it is low due to the relatively small masses of most bodies and owing to the large distances between them. The same is valid for the influence that moons have on the densities of their primaries. For another thing, primaries formed before moons, and, consequently, a primary influences its moon's density, rather than vice versa.

The mean density of a celestial body cannot be measured directly, but is always calculated, and the methods of the mean density determination do not differ fundamentally in this aspect, including the new approach proposed in the given article. However, to avoid uncertainty in terminology and mathematical notations, all values of the mean densities, obtained using traditional gravitational and non-gravitational methods, are referred to as the "observed" mean density and are denoted as $\rho^o$; the calculated values of the mean densities, by the method proposed in the given article, are referred to as the "calculated" mean density and are denoted as $\rho^c$.

## 3. Spatial distribution of the mean densities of major planets

Major planets are the largest and the most massive bodies in the Solar System along with the Sun, and therefore variation in their mean densities is to be considered first of all. The present article focuses on spatial distribution of the mean densities of celestial objects rather than on the mean density values as such. The spatial distribution of the mean densities of major planets is shown in Fig. 1.

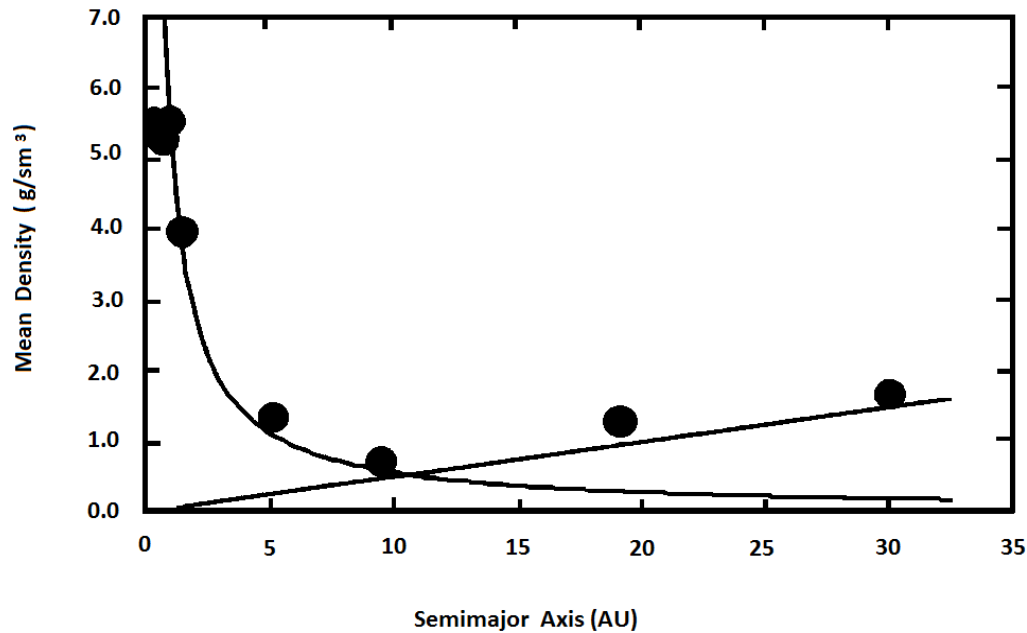

**Fig. 1.** The mean densities of major planets in the Solar System.

On looking at Fig. 1, non-randomness of the pattern salutes the eye. Instead of the chaotic character distribution of the mean densities, which should seem to happen, it is clearly seen in Fig. 1 that the mean densities change successively. The planets, namely Mercury, Venus and Earth, occurring at the nearest distances to the Sun, have the maximal mean densities. The planets Earth, Mars, Jupiter and Saturn have asymptotically decreasing values of the mean densities. The planets Saturn, Uranus and Neptune have the monotonously increasing mean densities. On this basis, it is assumable that distribution of the mean densities of the Solar System's bodies follows a certain regular pattern.

The first three planets, closest to the Sun, have almost equal mean densities (Fig. 1), and the change in the mean density of a planet versus the planet's distance to the Sun can only be described for the planets from Earth to Neptune. It is certainly imprecise to draw any adequate conclusion based on the analysis of the six bodies; however a provisional analysis is possible, considering the high accuracy of the mean densities of these planets.

The mean density of a planet can be expressed as a function of the planet's distance to the Sun:

$$\rho_i^c = f\left(R_i\right) + q \tag{1}$$

where $\rho_i^c$ is the "calculated" mean density of the $i$-th planet; $R_i$ is the semimajor axis of the $i$-th planet orbit; $q$ is a certain coefficient.

The observed change in the mean densities can be assumed as the result of influence of two additive factors, and can be presented as two differently directed curves (Fig. 1). In this case, each planet's mean density has two components, and grouping of the planets into two divisions is based on which of the components governs the dominant part of the mean density value. Each curve only characterizes a part of the body's mean density; consequently, each curve should be plotted below the mean density points. In the first approximation, the "descending" curve can be described by the inverse proportion to R; and the "ascending" curve — by the direct proportion to $R$. Let coefficient $q$ be zero, then Eq. (1) transforms into:

$$\rho_i^c\left(R_i\right) = \frac{k_1}{R_i} + k_2 R_i \tag{2}$$

Empirical coefficients are $k_1 \approx 5.6$ AUg/sm³ = $8.4 \times 10^{14}$ кг/м², $k_2 \approx 0.05$ g/AUsm³ = $3.3 \times 10^{-10}$ кг/м⁴.

The change in the mean densities of the planets according to the ratio $1/R$ suggests that this change depends on the gravitational (or inertial) potential, since the latter obeys the same ratio in case of the point sources. Thus, formation of the mean density of a body at a certain point of the Solar System could be conditioned by the value of the gravitational potential at that point. Inasmuch as the mean densities of the planets change as the functions of the planets distance to the Sun and considering that the Sun is the most massive object in the Solar System, it is possible to state that the mean densities of the planets were mainly affected by the Sun's gravitational field. But there are two members in Eq. (2) and two curves (Fig. 1). This implies that the mean densities could form under influence of two gravitating objects. This possibility will be discussed in Section 6.

## 4. Spatial distribution of the mean densities for major planets, dwarf planets, natural satellites and small Solar System objects

The relationship expressed by Eq. (2) should not only be valid for major planets, but it should hold true for other objects in the Solar System. This means that Eq. (2) should allow calculating the mean densities for dwarf planets and SSSBs for which is as well assumable that the Sun is the main gravitating object.

The situation is more complex in case of a planet's moons. Logically, the mean density of a moon formed under the gravitational influence of the Sun and the primary. The primary and its moons occur, upon average, at the same distance to the Sun, so, the Sun's gravitational field contributes equally to the mean densities both of an $i$-th planet, and each moon of this planet, $\rho_i$. Since quantities are assumed additive, the mean density of an $n$-th satellite of an $i$-th planet, $\rho_{sin}$, is found from the equation:

$$\rho_{\sin}=\rho_i+\rho_{in} \tag{3}$$

where $\rho_{in}$ is the contribution of the primary to the density of the $n$-th satellite. Accordingly, a part of the mean density, which formed under influence of the primary, is equal to the difference between the planet's satellite mean density and the planet's mean density: $\rho_{in}=\rho_{sin}-\rho_i$. This inference can apparently be related to both the "observed" and "calculated" mean densities, i.e.

$$\rho_{\sin}^{o}=\rho_i^{o}+\rho_{in}^{o} \tag{4}$$

$$\rho_{\sin}^{c}=\rho_i^{c}+\rho_{in}^{c} \tag{5}$$

The integrated analysis of the densities of all celestial bodies supposes studying the effect of the central gravitating object. This means estimation of influence exerted by the Sun on major planets, dwarf planets and small Solar System bodies, and the influence exerted by primaries on their natural satellites. Consequently, the primary-affected part of the mean density of a satellite, $\rho_{in}$, should only be considered. Inasmuch as it is supposed that gravitational influence, conditioned by gravitational potential, is governed by the mass of the central body for the planets and the planet satellites, then the effect of the Sun on all celestial bodies and the effect of a planet on its moons are proportional to the ratio of masses of the Sun and that planet. Assuming the mean density formation law uniform, Eq. (2) should keep its form, but the coefficients $k_1$ and $k_2$ for a planet should change in proportion to the mentioned ratio: $k_1$ should decrease and $k_2$ should increase. On the other hand, each term in Eq. (2) includes $R$ to the first power. This allows a conditional assumption that the coefficients $k_1$ and $k_2$ remain unaltered for all gravitating objects, but the distance scale is what changed; then in the distance scale, the mean distance between a planet and its satellites, $R_{in}$, may be given as $r_{in}$ in relative astronomical units (rAU); i.e.: $r_{in} = R_{in} \times M/m_i$, where $M$ is the solar mass, $m_i$ is the mass of $i$–th planet.

The relationship between the "observed" values of the mean densities of the major planets, dwarf planets and SSSBs $\rho_i^{o}$, planet's satellites $\rho_{in}^{o}$, and their distances to the central body ($R_i$ or $r_{in}$) is depicted in Fig. 2.

The picture shows the mean densities of several kinds of objects: there are 8 major planets, 4 dwarf planets, 31 natural planet satellites, as well as small Solar System bodies, including 70 asteroids, 7 TNOs and 11 comet nuclei. The dwarf planets are represented by Ceres, located in the main asteroid belt, and three trans-Neptunian objects: Pluto, Haumea and Eris. Asteroids make up most of the SSSBs; these are 7 near-Earth asteroids, one Trojan asteroid Patroclus and the rest are the main belt asteroids. There are 16 Saturn's natural satellites, 5 Jupiter's satellites, 5 satellites of the Uranus. Besides, the Earth, Neptune and Pluto have one satellite each. All in all, the mean densities of 131 celestial objects are presented in the Fig. 2.

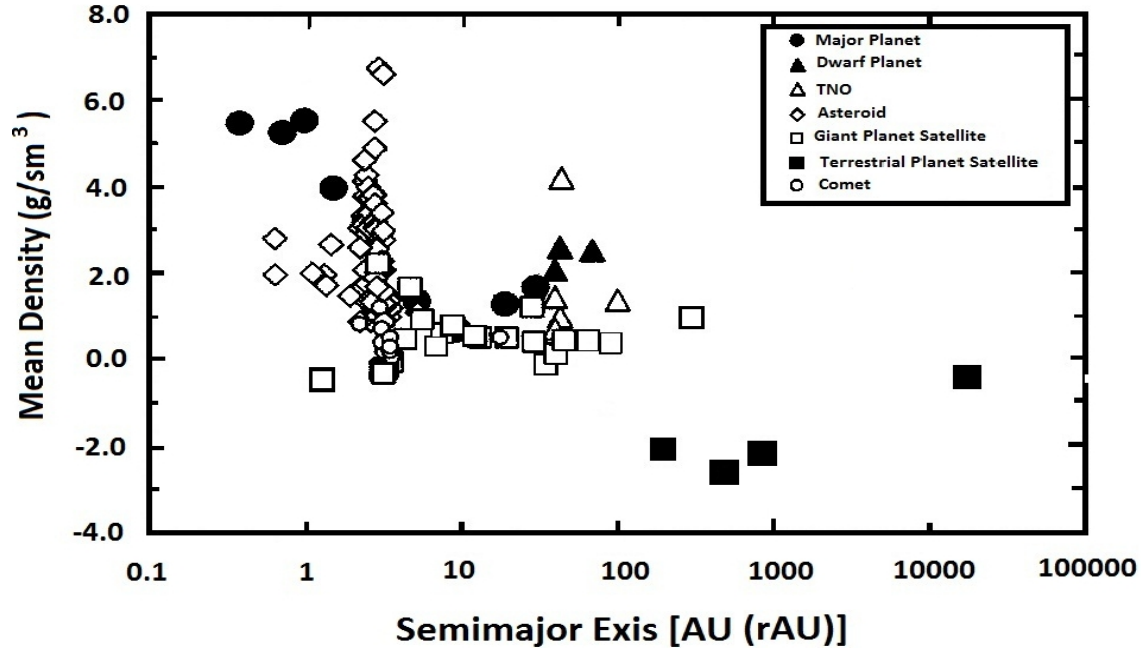

**Fig. 2.** Spatial Distribution of the Mean Densities for the Solar System bodies.

The mean densities of all objects, shown in Fig. 2, do not provide a clear picture of change in the mean density value versus the object's distance to the Sun, though there are a few partial dependences.

The three dwarf planets in the trans-Neptunian region follow an upward trend in the mean density value with the distance to the Sun in compliance with the "ascending" curve of the mean densities of the major planets in Fig. 1. The other trans-Neptunian objects weakly exhibit the said trend.

The main belt asteroids are concentrated within a relatively narrow value range of the distances to the Sun (Fig. 2); therefore their mean density trend is subtle. We allocate the asteroids into two groups, according to the values of the mean density. The first group includes a few asteroids with their mean densities higher than 4 g/cm$^3$. The mean densities of these asteroids show the upward trend with the distance to the Sun, though that inference is rather weak inasmuch as there are few objects to make the sampling sufficient. The second group consists of the asteroids with their mean densities below the value of 4 g/cm$^3$, and these asteroids show the downward trend with the distance to the Sun. The latter trend is weak, and is smoothed by a logarithmic scale, but it includes nearly 60 asteroids. Most comets exhibit the same relationship.

More than 10 natural satellites at a distance from their primaries, ranged from 3 to 100 relative astronomical units, have the downward trend in the mean density as the semimajor axis of the orbit grows (Fig. 2). Pluto's moon Charon, shown here as a terrestrial planet satellite, above 10,000 rAU apart, as if extends the trend, but its mean density is a negative value. Less than 3 relative astronomical units off the Sun, the mentioned trend shows itself clearly, and there are two alternatives of the trend. The first alternative is the intensive increase in the mean density value on approach to the Sun, which is a sort of the "descending" curve in Fig. 1. The second alternative suggests that these natural satellites make a united group with the near-Earth asteroids; in the latter case, the mean densities of the bodies grow on approach to the Sun, too, but less intensively.

A group of the natural satellites of the terrestrial planets, i.e. Moon, Phobos and Deimos, have very low, negative mean densities — nearly -2 g/cm³ (Fig. 2). Some other satellites are also characterized by the negative mean densities. Evidently, these negative values of the mean densities are the result of the calculation, have no physical sense and require an individual analysis. Probably, a few of the negative values, which are slightly below zero, may be the result of the calculation inaccuracy, while anomalously low, negative-valued mean densities have another source to ensue from. An explanatory suggestion will be given in Section 6 of this article.

To sum up the discussion of Fig. 2, there is no a well-defined dependence of the mean densities of bodies and the distances of the bodies to the Sun.

## 5. Spatial distribution of the mean densities of massive bodies in the Solar System

The above analysis shows that there is a regular pattern in the spatial distribution of the mean densities of major planets (Fig. 1) and there is no regular pattern in the spatial distribution of the mean densities for the entire set of celestial objects (Fig. 2). Thus, Eq. (2) and Eq. (4) are only usable to characterize a part of the objects that are supposedly the most massive bodies in the Solar System. These are major planets, dwarf planets and the most massive planet satellites. Spatial distribution of the mean densities of the said celestial objects is shown in Fig. 3.

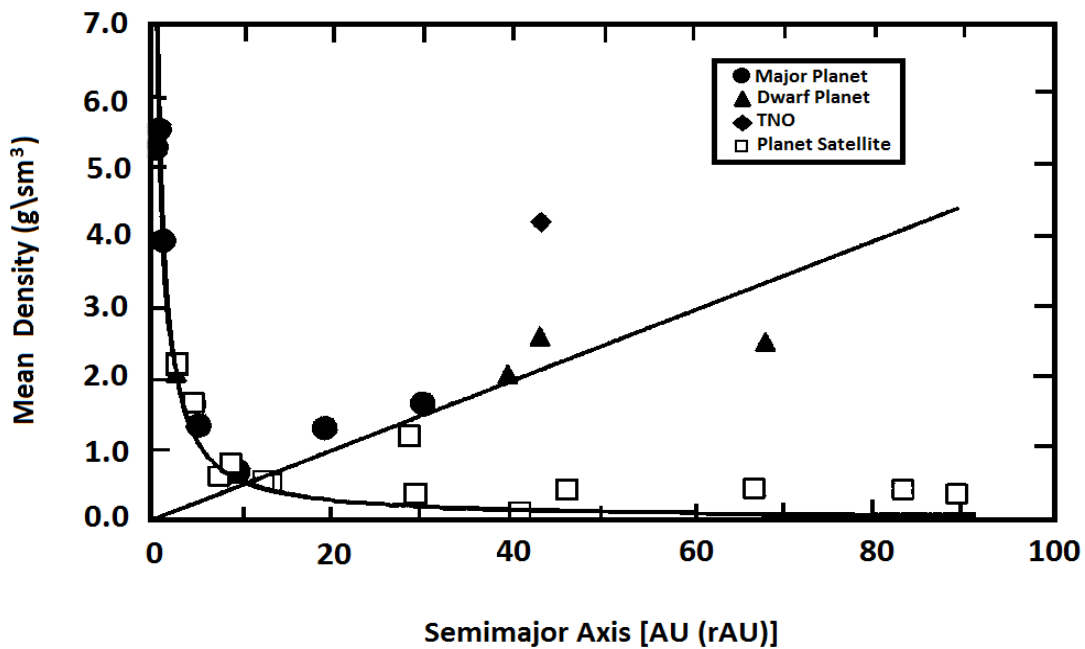

**Fig. 3. S**patial distribution of the mean densities of massive objects in the Solar System.

Figure 3 displays the same two trends in the mean density distribution of the bodies as in Fig. 1. The first trend is the gradual decrease in the mean densities with the increasing distance to the Sun or the primary; the second trend is the gradual increase. The first trend characterized by the "descending" curve describes the mean densities of the planets from Earth to Saturn, and includes the dwarf planet Ceres as well as 12 big moons of major planets. Ceres, Io, Europa, Ganymede, Dione complement the curve, and Callisto, Rhea, Iapetus, Ariel, Umbriel, Titania, Oberon, and Triton extend it. The second trend characterized by the "ascending" curve illustrates the increase in the mean densities of the bodies in the direction off the Sun. These are the mean densities of the planets from Saturn to Neptune and of three dwarf planers. Pluto's mean density quite fits the curve. The other massive dwarf planers, namely, Haumea and Eris, have high values of the mean density, which reckons them among the bodies belonging to the second curve; however these mean densities largely deviate from this curve.

Three massive natural satellites seem abnormal: these are the Earth's satellite Moon and Pluto's satellite Charon (Fig. 2), and Saturn's satellite Titan (Fig. 3). The Moon and Charon have negative-valued mean densities. They are not shown in Fig. 3 as they are situated very far from their primaries. Titan stands out of the relationship obtained for satellites, and is more like inclined to the ascending curve.

It is interesting to obtain the lower estimate of the planetary mass range, at which the mean density of a planet fits the revealed relationship. The least massive object with its mean density being in accord with the obtained relationship is dwarf planet Ceres. The largest objects, though their masses are, evidently, insufficient to satisfy the relationship, are TNOs Orcus and Varuna and Saturn's satellite Tethys. The comparison of the masses of Orcus (Brown et al., 2010), Varuna (Lacerda et al., 2007), Tethys (Thomas, 2010), on the one hand, and Ceres (Baer et al., 2011), on the other, allows judging that the lower estimate of the mass for the objects, to be in agreement with the revealed relationship, is within $7$-$9 \times 10^{20}$ kg. The objects with the masses above the indicated estimate will be referred to as "massive".

The "observed" mean densities of major and dwarf planets, and TNOs, $\rho_i^o$, as well as the "observed" mean densities of satellites, $\rho_{sin}^o$, are determined by the traditional gravitational and non-gravitational methods. The respective "calculated" mean densities $\rho_i^c$ and $\rho_{sin}^c$ are determined using the method proposed in the given article. The "calculated" mean density value includes two components that differ in the estimates for planets and for satellites. These components for major and dwarf planets and for TNOs correspond to the terms in Eq. (2). The components for satellites follow from Eq. (5). Table 1 presents the difference of the "observed" and "calculated" values of the mean densities, while the root-mean-square deviation (RMSD) of them is given in Table 2.

Table 2 shows the root-mean-square deviation of the "calculated" mean densities relative to the "observed" mean densities of six major planets equals approximately 0.20 g/cm$^2$. For the analyzed set of 26 bodies, the RMSD makes up 0.68 g/cm$^3$. We have already noticed earlier that the mean densities of Mercury and Venus, calculated by Eq. (2), are off the actual values, the negative mean density of the Moon is anomalous as well. In addition, the mean densities of Quaoar, Eris and Titan should most probably be assumed as anomalous. A substantial deviation features the mean density values of Europa and Charon. Thus, we have that at least 6-8 in 28 massive bodies in the Solar System, or one fourth of the total amount that has been analyzed, have an abnormal (divergent) mean density value; the RMSD for the rest of the bodies is approximately 0.22 g/cm$^3$.

**Table 1**

The difference between the "observed" and "calculated" mean densities of massive bodies in the Solar System.

| Name | Difference (g/sm$^3$) | Name | Difference (g/sm$^3$) |
|---|---|---|---|
| Mercury* | — | Io | -0.30 |
| Venus* | — | Europa* | -0.46 |
| Earth | 0.13 | Ganymede | 0.14 |
| Mars | -0.18 | Callisto | -0.07 |
| Jupiter | 0.01 | Dione | 0.22 |
| Saturn | 0.38 | Rhea | 0.28 |
| Uranus | -0.02 | Titan* | -0.62 |
| Neptune | 0.05 | Iapetus | 0.04 |
| Ceres | 0.07 | Ariel | -0.22 |
| Pluto | 0.05 | Umbriel | 0.00 |
| Haumea | -0.32 | Titania | -0.37 |
| Quaoar* | -1.91 | Oberon | -0.32 |
| Eris* | 0.97 | Triton | -0.25 |
| Moon* | 2.31 | Charon* | 0.48 |

* "Anomalous" object

**Table 2**

The root-mean-square deviation between the "observed" and "calculated" mean densities of massive bodies in the Solar System.

| Quantity of bodies | Deviation (g/sm$^3$) |
|---|---|
| 6 Major Planets | ±0.20 |
| 26 Bodies | ±0.68 |
| 20 Bodies | ±0.22 |

Let us determine an average error of assessing the mean densities of celestial bodies by the gravitational and non-gravitational methods. It will be observed that the mean density accuracy is very different for different bodies. For the most massive objects, such as the major planets and the biggest natural satellites, the mean densities are estimated accurate within four or five significant digits, while small or remote objects have their mean densities assessed to the accuracy of two or one significant digits. Figure 4 presents the mean density errors for the bodies in the Solar System.

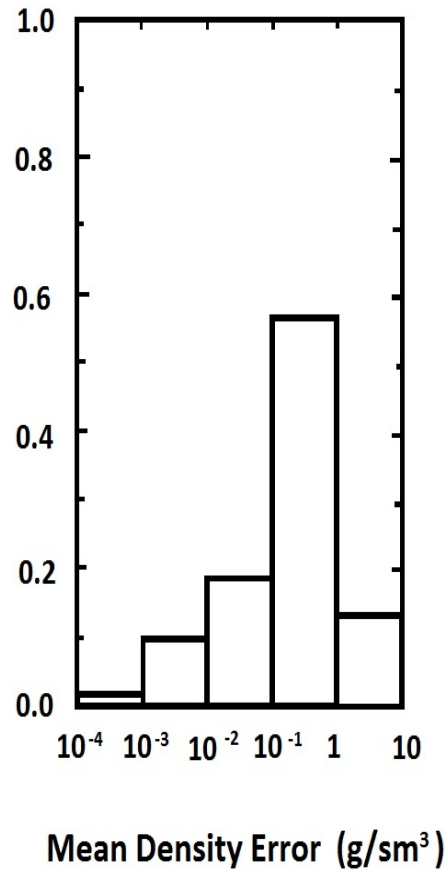

**Fig. 4.** The error histogram for estimates of the mean densities of the bodies in the Solar System.

The histogram shows that more than a half of the mean density errors fall within 0.1-1.0 g/cm³, with a modal value about 0.3 g/cm³. Table 2 above in the article shows the RMSD in estimating the mean densities by the newly proposed method, as compared with the traditionally determined values, equals approximately 0.2 g/cm³ for the many of massive bodies. Hence it appears that the estimates of the mean densities by the traditional methods and the new approach exhibit quite comparable accuracy. Thus, it is reputed that the mean densities of the majority of massive bodies in the Solar System are possible to highly accurately calculate only using the data on their distances to the Sun, or, for satellites, the distances to the Sun and the primary.

## 6. Discussion and hypotheses

According to the discussed analysis, the mean densities of major and dwarf planets, as well as of massive natural satellites are governed by the location of these objects in the Solar System. So, based on Eq. (2), the mean densities of the Solar System bodies form a continuum where each point corresponds to a certain density. This allows a supposition on the existing spatial distribution of the mean densities of the Solar System bodies. Consequently, the bodies, that are thousand times and million times different in mass and size, with different internal structures and chemical compositions, have mean densities that depend on the body's distance to the Sun, or in case of natural satellites—on the satellite's distance to its primary, as well. Based on that, we deduce an inference on the existence of the natural regulatory mechanisms for the mean densities of celestial bodies, independent of other physical and chemical characteristics of the bodies. Moreover, it is possible to assume some of these characteristics secondary relative to the mean density; that is, the mean density value governs these characteristics rather than vice versa. For example, a planet with a definite mass could not have an "arbitrary" volume, as the volume is governed by the density that suits the gravitational potential at the given point of the Solar System. What has been said seems extraordinary, and yet, the revealed relationship is hard to explain elsewise.

A question arises: why is this relationship rather clearly evident for the massive objects only? To all appearance, the bodies, which obey the revealed mechanism, must be well-shaped and have sufficient mass to achieve hydrostatic equilibrium, which shows itself in their spherical form. However, not all of the bodies abide by the revealed relationship; for some massive bodies, the mean densities obtained by the traditional methods differ greatly from the mean densities found using the new proposed method (Table 1). Naturally, the deviation may ensue from the determination uncertainty of both the traditional and new approaches. At the same time, another fact is more important: it is always meant that any error is determined relative to a true value. But in our case, some, and even most of deviation between "observation" and "calculation" mean density values are probably governed by natural reasons, rather than by the calculation and determination uncertainties; therefore the deviation in the values cannot be assumed "errors". We may hypothesize that while a body is forming, the body's mean density value is fixated and never changes appreciably if no disastrous events (giant impacts) take place. Consequently, this article offers the method to determine a body's mean density value coincident with the moment of the body formation and with the formation conditions. In addition, the determined mean density shows the initial position of a body in the Solar System. In case that a body did not form in situ but was transferred to the given position after it had formed, the body's mean density value will not match the value calculated by Eq. (2). The transference means the change in parameters of the body's orbit rather than in the orbital motion. In other words, the bodies that changed their distance to the Sun as well as the natural satellites that changed their distance to the primary have other mean density values as are calculated by (2). And, as we have already said, a disastrous event can change the mean density of a body. In both cases, the calculated density will appear as an outlier.

It is an interesting fact that most satellites of large planets have higher mean densities as compared to their primaries, while satellites of terrestrial planets and Pluto's satellite have lower mean densities as compared to their primaries. This is well seen in Fig. 2 where the illustrated mean densities of satellites, $r^o_{in}$, are the results of the calculation by Eq. (3). The values shown in Fig. 2 represent the difference between the mean densities of the primary and its satellite; therefore, the positive valued mean densities of majority of the satellites imply the satellites possess the higher mean densities than their primaries. Alongside with that, the mean densities of the terrestrial planets' satellites and Pluto's satellite, $r^o_{in}$, have negative values. A straightway hypothesis is that large planets have low and easy-to-overtop mean densities, and the terrestrial planets have high and difficult-to-surpass mean densities. Although tenable for the terrestrial planets, this hypothesis is unadoptable for giant planets such as Jupiter, Saturn, Uranus or Neptune. The huge gravitation of giant planets in the course of their formation should result in that heavy elements were attracted and merged into the giants rather than were left in circumplanetary discs which became the source for the satellites of these giant planets to be formed later on. For another thing, pressure inside a planet, dependent on the planet's mass, should be higher than the pressure inside the planet's satellites, which also should contribute to the higher mean density of the planet as against the planet's satellites.

It is worthy of mentioning that most major planets have the higher mean density values than the Sun, which implies an analogous relationship of the massive center and its satellites in the framework of the entire Solar System. And this highlights once again the existence of the specific natural mechanisms of the mean density formation.

Using the described concepts of the mean density formation as the basis, it is possible to get an insight into causes of the revealed discrepancies and to give a reason for the abnormal mean densities of the concrete objects. So, all terrestrial planets' satellites, Moon, Phobos and Deimos, as well Pluto's satellite, Charon, possess lower value mean densities as against the mean densities of their primaries (Fig. 2).

Two of the listed satellites, Moon and Charon, are massive and should obey the revealed relationship, one would think. In this context, it gets into the field of attention that the listed satellites occur much farther from their primaries as compared to the other major and dwarf planets, SSSBs and natural satellites of large planets (Fig. 2); or, to put it more exactly, the listed satellites fall in the area of the appreciably lower gravitational potential. This makes it allowably to hypothesize that the primaries captured these natural satellites. The most adequate current hypothesis suggests the impact origin of the Moon (Cameron and Ward, 1976; Benz et al., 1986; Cameron, 1997; Canup, 2004; Canup and Barr, 2010), Charon (Tholen and Buie, 1997; Canup, 2005, 2011), Phobos and Deimos (Craddock, 2011).

One way or another, the thing is that the Moon and Charon had not formed under the influence of gravity fields of their primaries, which is an explanation of their abnormality, in a sense. Almost every object of the main asteroid belt dissatisfies the evolutionary formation condition, as these objects, except for Ceres and Vesta, are not assumed as the independently formed bodies but the fragments of larger bodies (Murray and Dermott, 1999). What has been said above relates to most of the Kuiper Belt objects, the impact history of which throws back, thus, it is possible to regard them victims of disastrous events that changed their mean densities (Davis and Farinella, 1997, Fraser, 2009, O'Brien et al., 2005, Charnoz, 2010). For instance, Haumea is, apparently, a fragment of a larger predecessor (Levison et al., 2008). In addition, many Kuiper Belt objects could be displaced after formation, either due to collisions or under the influence of a hypothetical planet that migrated afterwards (Emel`yanenko, 2010).

Another exception is Saturn's moon, Titan, the only one to exhibit abnormality among 12 massive moons of major planets. The discrepancy between the "observed" and "calculated" values of Titan's mean density is impossible to explain by the specific character of Saturn, since the other massive moons of this planet, including remote Iapetus, satisfy the discussed relationship. An opinion on Titan to be formed otherwhere and then to be trapped by Saturn is advanced by Prentice (2007), which is a rather rare idea, as Titan has a nearly circular orbit, and such orbit is, as a rule, assumed the satisfactory argument in favor of the in situ formation of the satellite. At the same time, if we took the value of Titan's mean density and calculated initial position of Titan by Eq. (2), the result would not seam arbitrary. To a high precision (0.04 AU), the calculated initial position of Titan would fit a 2:1 resonance with Jupiter, and Titan would lie at outer boundary of the main asteroid belt in relation to the Sun. Formation of Titan in the to-be Kirkwood gap could destabilize the orbit, and, as a consequence, Titan could be thrown to Saturn orbit. In the described scenario of Titan's initial position, the size and mass of Titan would allow assuming it the 9th, more properly, the 5th major planet. It is also worth saying that Titan is the only satellite in the Solar System to possess the stable atmosphere (Niemann et al., 2005, Owen and Niemann, 2009, Dorofeeva and Ruskol, 2010) and, probably, life (Strobel, 2010). Both facts are atypical for a satellite.

As for Mercury and Venus, it is suitable to suppose that the mean density values undergo natural constraints in the Solar System. The calculated mean densities of Venus and Mercury by Eq. (2) are 7.7 and 14.4 g/cm$^3$, respectively. It is probable that the protoplanetary disk lacked the sufficient amount of heavy elements, and the gravity forces failed to ensure the required internal pressure, and the formed body did not acquire the "desired" mean density. The existing specific physicochemical conditions governed the mean density limit in the given area of the Solar System, and that predefined the impossibility for the mean density value to overcome the local limit of 5.2-5.5 g/cm$^3$. As like as not, this explanation may be valid for the natural satellites of Earth and Pluto.

Massive bodies are certainly less movable under collisions or resonance alignments of positions in space. According to Table 1, the "anomalous" bodies make up one fourth of the total set of the massive bodies in the Solar System. This means that orbital distances of the most massive objects have not changed, which enables a definitively reliable estimate of the mean densities of massive TNOs by Eq. (2). For example, Pluto's mean density values found by the traditional approach and the proposed method are nearly equal (Table 1). The mean densities of Haumea and Quaoar could make up 2.24-2.25 g/cm$^3$. Let us estimate the mean densities of the bodies, the mean densities of which have not been determined yet: dwarf planet Makemake should have the mean density approximately 2.35 g/cm$^3$; dwarf planet Eris and object 2007 OR10—nearly 3.4 g/cm$^3$. It is though not known for sure so far if Eq. (2) is applicable to extrapolation to the distances where the latter objects are, since the said equation can be imprecise.

The first term of Eq. (2) shows an inverse proportionality between the mean density of a planet, and the planet's distance to the Sun. This can be assumed an accurate value as most objects, which are discussed in this article, can be taken as points. The second term of Eq. (2) keeps us guessing why the values of mean densities of some objects increase with the increasing distance to the Sun. Accordingly, a precise mathematic formula of this term is impossible to derive.

Let us compare the Sun—planets system and the planets—satellites system. The distribution of the mean density in the former system is vitally different from the mean density distribution in the latter, which is illustrated by the two curves in Fig. 3. The ascending mean density curve does only exist in the system of the Sun and major and dwarf planets, and is absent in the systems of planets and their satellites. The term $k_2R$ of Eq. (2) shows itself in the Solar System and does not occur in the planets-satellites system. It seems the most reasonable to explain the increment in the mean densities of the remote planets in the Sun—planets system by the presence of a hypothetical object as an additional source of gravity. According to the new hypothesis that is discussed in the given article, the mean density of an object is governed by the gravitational potential that is conditioned by the gravitating masses; accordingly, the values of the mean densities on the ascending curve indicate that a hypothetical object has immense mass, commensurable with the mass of the Sun.

On the other hand, the ascending curve describes the central symmetry of the system. This means the distribution of the mean densities is conditioned by the Solar System bodies rather than by remote objects. For instance, the Galaxy Center would equally contribute to the mean densities of all objects of the Solar System. Moreover, considering the direction of increment in the mean density values, that hypothetical object should occur in the periphery of the Solar System rather than in its center. Consequently, a hypothetical object possesses the following characteristics:

1. It is the object of the Solar System;

2. It has huge mass;

3. It is symmetric relative to the center of the Solar System;

4. It occurs in the region beyond Neptune.

Based on the assembly of the listed characteristics, such object could be a gravitating sphere or a gravitating belt in the trans-Neptunian Solar System. However, physically, neither a sphere nor a belt in the ecliptic plane can be such an object since their inner gravitational field is equipotential. This object may be, for instance, a belt situated at an angle to the ecliptic and, possibly, orthogonally to it. That belt could contain a variety of relatively small bodies. But in order to possess a sufficient gravitational potential, this hypothetical belt, even if situated at a distance at which the current Kuiper belt is, should have a great mass. Most probably, such object existed during the period of formation of planets and had lost its mass by now. For instance, Stern and Colwell (1997) determined a big mass of the primordial Edgeworth-Kuiper belt. The deficit and inaccuracy of definitions existing for the mean densities of the known remote TNOs makes it difficult to determine possible location and characteristics of this hypothetical belt.

The traditional methods, on the one hand, and the proposed approach, on the other hand, enable independent estimates of one and same property of celestial objects, namely, the mean density, but at different life stages. The comparative analysis of the data obtained by different methods is suitable to use in researching migration of objects in the Solar System and in other planetary systems.